# Neutron-Gamma Pulse Shape Discrimination for Organic Scintillation Detector using 2D CNN based Image Classification.

**Annesha Karmakar[1,3], Anikesh Pal[2], G. Anil Kumar[3*], Bhavika[3], V. Anand[3], Mohit Tyagi[4]**

[1]Nuclear Engineering and Technology Program, Indian Institute of Technology Kanpur, India
[2]Department of Mechanical Engineering, Indian Institute of Technology Kanpur, India
[3]Radiation Detectors and Spectroscopy Laboratory, Department of Physics, Indian Institute of Technology Roorkee, India
[4]Technical Physics Division, Bhabha Atomic Research Centre, Mumbai, India

*anil.gourishetty@ph.iitr.ac.in

## ABSTRACT

This study shows an implementation of neutron-gamma pulse shape discrimination (PSD) using a two-dimensional convolutional neural network. The inputs to the network are snapshots of the unprocessed, digitized signals from a BC501A detector. By exposing a BC501A detector to a Cf-252 source, neutron and gamma signals were collected to create a training dataset. The realistic datasets were created using a data-driven approach for labeling the digitized signals, having classified snapshots of neutron and gamma pulses. Our algorithm was able to successfully differentiate neutrons and gammas with similar accuracy as the CI approach. Additionally, the independent dataset accuracy for our suggested 2D CNN-based PSD approach is 99%. In contrast to the traditional charge integration method, our suggested algorithm with data augmentation, is capable of extracting features from snapshots of the raw data based on the signal structures, making it computationally more efficient and also appropriate for other types of neutron detectors.

## Introduction

Pulse shape discrimination (PSD) techniques are necessary for nuclear radiation studies for a variety of purposes, including nuclear security and safeguard applications, medical imaging, and spectroscopy. The PSD is a common categorization problem in which the pulses are distinguished based on their shapes. One of the typical characteristics of PSD is, for instance, the decay time of detected scintillator light. The various paths of transitions from excited states to ground state in the scintillation detector medium determine this characteristic. Traditional digital PSD methods like charge integration (CI) method[1-4](and references therein) are popular for their robust and reliable performance[5], but are dependent on a variety of factors like scintillation material choice, experimental techniques, signal readout techniques, etc. Standard digital PSD techniques have limitations in using real-time data that exhibit a number of features such as pulse pile-up and background interactions. Additionally, traditional PSD methods are unable to handle multiple radiations simultaneously. Machine learning, especially Neural Networks have superior classification capability, provided they are trained with realistic data. While previous studies reported the use of machine learning algorithms for PSD[6-10], it is the Convolution Neural Network (CNN) which is very well suited for raw data that have high presence of local correlations i.e. different properties of waveform signals. As CNNs are used in computer vision tasks (image processing), it is capable of reaching high performance with a limited number of datasets. Once properly trained, the discrimination capability of a CNN is also very fast. There have been several applications of CNNs in image classification, especially in the medical imaging field for the identification of diseases. The examples are breast cancer classification[11], Glioma classification[12], neck and head cancer classification[13], grading of fatty liver from segmented images of sonograms[14] etc. Before training with limited data sets, the image classification algorithms addressed in the literature involve pre-processing of the image, such as cropping, choosing the region of interest for the image, or segmenting the image. There have been a few recent studies where the identification of liver cancer, its preliminary and secondary grading is done from raw 3D tomographic MRI images[15] and automatic identification of breast cancers from whole mammogram images, without the requirement of initial pathological detection, segmentation or image cropping. Identification and classification of objects which require expertise is possible using the CNN method of image classification. Therefore, this technique can be applied to pulse shape discrimination (PSD) of neutrons and gamma rays using raw pulses. Although PSD is a very common measure of the performance of scintillation detectors in mixed radiation fields, the application of image processing techniques in the genre of nuclear radiation detection is very limited. A recent study reported the discrimination capability of 1D CNN for electron signals and nuclear signals for $^6LiF:ZnS$(Ag) detector and compared it with CI method, in order to provide a benchmark for the discriminating capability[16]. This study has demonstrated high-level performance of discrimination for electron and neutron signals with an AUC (Area Under the Curve) of 0.996+_0.003 when compared to the traditional method. However, this study is only limited to the performance of a single detector. In this study, we present the neutron gamma discrimination capability of two-dimensional convolution neural network (2D CNN) for two different source-detector combinations in mixed radiation fields.

      To the best of our knowledge, for the first time, we have used the technique of image processing in the pulse shape discrimination method and benchmarked its performance to the traditional CI method. This image processing algorithm is trained using the snapshots of the raw pulses of neutrons and gammas from BC501A using a Cf-252 source. This detector is commonly used for both neutron and gamma detection and inherently possesses good neutron-gamma PSD capability. The architecture pipeline consists of three stages, namely, data extraction for training, data augmentation and the 2D CNN model. In order to ensure its viability and correctness, our proposed algorithm is also tested on an independent dataset generated from a different source-detector combination (BC501 and Am-Be source). The discrimination performance of our algorithm is presented and discussed in the form of a confusion matrix.

## Methods

### 0.1 Experiment for training dataset

For obtaining neutron and gamma pulses, an experiment was conducted that can discriminate neutrons and gammas using CI method. A BC501A organic liquid scintillation detector is situated 5 cm away from a 10 $\mu Ci$ Cf-252 source. The output of the photomultiplier tube (PMT) of the detector is connected to a 1 GHz CEAN DT-5751 desktop digitizer with an integrated multi-parameter spectroscopy program called "CoMPASS" at a sampling rate of 1 GS/s to record the neutron and gamma pulses[17]. This experiment is performed at Radiation Detectors and Spectroscopy Laboratory, IIT Roorkee, India.

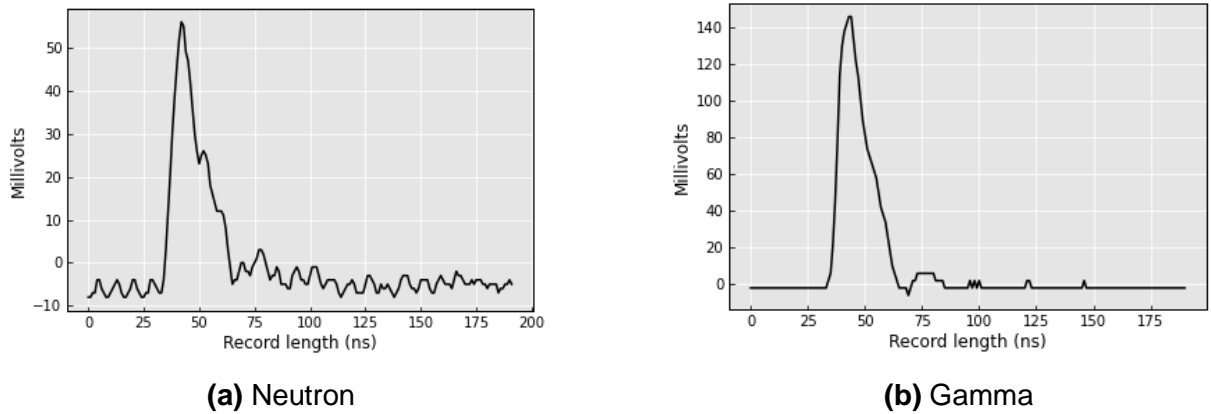

(a) Neutron  (b) Gamma

**Figure 1.** Sample snapshots of input data.

### 0.2 Experiment for independent dataset

A 30 mCi Am-Be source is positioned 6 cm away from the front of a BC501 cylindrical detector with dimensions of 5.08 cm radius and 1 cm length. A 0.25 GHz DT 5790 digitizer is used to acquire neutron and gamma pulses. This experiment is conducted at Technical Physics Division, BARC, Mumbai, India.

### 0.3 Data extraction for training

The classified signals from the digitizer are extracted in the form of arrays where data points are stored to form the pulses. To obtain the snapshots, first, we normalize the pulses by subtracting each array from its minimum value. After normalization, a `for' loop is run from zero to the length of the number of arrays saving each plot in the .png format. These classified snapshots of the raw, digitized pulses, are used to train the CNN architecture for classifying neutrons and gammas as shown in figure 1.

**Table 1.** Parameters of layers of 2D CNN-PSD architecture

| Layers | Filter Size | Activation function |
|--------|-------------|---------------------|
| Conv | 32 | ReLU |
| Max-pool | 2 | - |
| Conv | 32 | ReLU |
| Max-pool | 2 | - |
| Conv | 64 | ReLU |
| Max-pool | 2 | - |
| Flatten | - | - |
| Dense | 128 | ReLU |

### 0.4 Data Augmentation

It is well known that the performance of the CNN algorithm depends on the size of the input data. For this purpose, we use a data augmentation technique which can enhance the size of the input dataset by generating a new dataset from the original input image. The size of the new dataset from each original training image is determined by the number of functions utilized in the data augmentation. There are a variety of methods for enhancing the image data[18] and in the present work we chose to magnify, rotate, and horizontally flip the image. This task can be carried out either before, after or during the training phase. In order to conserve disc space, we performed data augmentation in our algorithm during the training phase. The expanded dataset's images are also downsized to 224 x 224 pixels in size.

### 0.5 CNN architecture

The CNN architecture was created to categorize pulses into neutrons or gamma rays. The specific parameters of the convolution and pooling layers are given in table 1.

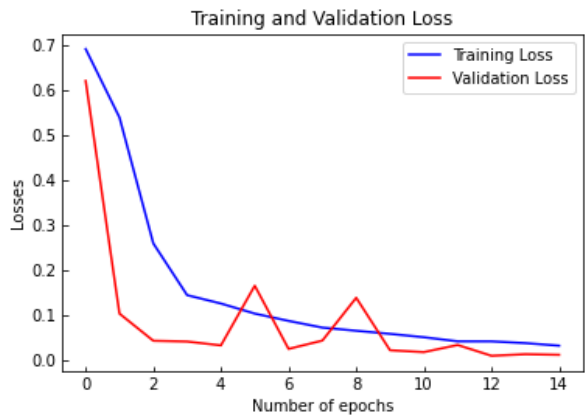

**Figure 2.** Comparison of training and validation loss of 2D CNN

Every signal snapshot is transformed by the CNN using three convolution and pooling layers, each using a non-linear ReLU activation functio[19]. A fully connected layer with 128 neurons that contain the ReLU activation function receives the output feature maps. The likelihood that the signal is neutron or gamma can be inferred from the output of the CNN, which consists of two neurons with a sigmoid activation function[20]. The Tensorflow[21] backend and Keras[22] were used to train the two-dimensional CNN. By decreasing the categorical cross-entropy loss on a training sample of 120,000 signals with an equal distribution of neutron and gamma signal snapshots, the network's weights were tuned. These signals were divided into training and validation data sets using the function "train_test_split" where the validation set comprises 30% of the data. The network was trained for a total of 15 epochs using the gradient descent (SGD) optimizer[23] at a learning rate of 0.01 with a batch size of 32 samples. The total time required for training using a 2.0GHz Intel $i$5 turbo boost system is approximately 1 hour and 4 minutes. The training and validation losses are shown in figure 2.

**Results**

A table that summarises how well a categorization algorithm performs is called a confusion matrix which is made up of four fundamental parameters (numbers) as given below.

1. TP (True Positive): TP represents the number of gammas that have been properly classified to be gammas.
2. TN (True Negative): TN represents the number of correctly classified neutrons.
3. FP (False Positive): FP represents the number of misclassified gammas but actually they are neutrons.
4. FN (False Negative): FN represents the number of misclassified neutrons but actually they are gammas.

Performance metrics of an algorithm are accuracy, precision, recall, and F1 score, which are calculated on the basis of the above-stated TP, TN, FP, and FN. The accuracy of an algorithm is represented as the ratio of correctly classified gammas and neutrons (TG+TN) to the total number of gammas and neutrons (TP+TN+FP+FN). The precision of neutrons ($PN$) or gammas ($PG$) is defined as the ratio of TN to the total number of predicted neutrons (TN+FN). Similarly, ($PG$) is represented as the ratio of TG to the total number of predicted gammas (TG+FG). The recall metric for neutrons ($RN$) or gammas ($PG$) is the ratio of the accurately classified neutrons or gammas to the total number of actual neutrons or gammas. The F1 score states the equilibrium between precision and recall by computing the harmonic mean of precision and recall.

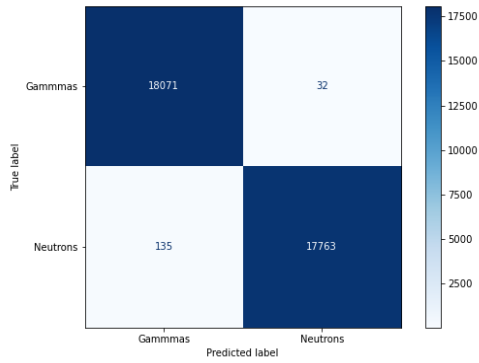 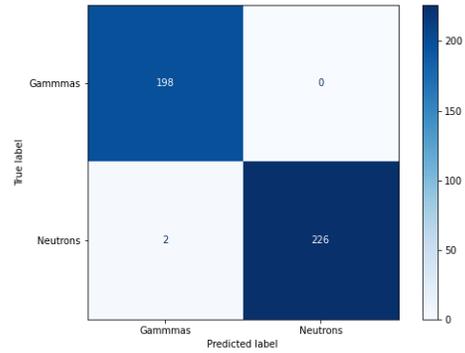

(a) Validation dataset                                            (b) Test dataset

**Figure 3.** Confusion Matrix for validation and test dataset.

The "True Label" in the confusion matrix is the classification of neutrons and gammas of training snapshots under consideration using the CI method and the "Predicted Label" represents the classification of the same using the 2D CNN-based PSD algorithm. The confusion matrix for the validation dataset is shown in figure 3a. The true neutron count is 135 + 17763 = 17898 and the true gamma count is 18071 + 32 = 18103. The predicted neutron count is 17763 and the predicted gamma count is 18071. Therefore, the precision for neutrons and gammas are 100% and 99% respectively. Corresponding recall values of neutrons and gammas are 99% and 100%. The F1-score of both gammas and neutrons is 100%. All the metrics indicate that our 2D CNN-PSD model predicts neutrons and gammas for the validation dataset with good accuracy and negligible false prediction. Further, we test the algorithm to a total of 426 pulses which was seen by the algorithm while training. These pulses are from the same detector and source combination and consist of 198 gammas and 228 neutrons. Figure 3b shows the confusion matrix of the test data set. It can be seen that the precision of gammas is 99% and that of neutrons is 100%. The corresponding recall value of neutrons and gammas are 99% and 100% respectively. It can be seen that the model can very accurately predict the neutron and gammas when compared to the CI method.

The efficiency of our proposed 2D CNN-PSD model along with the benchmark CI method was evaluated for an independent dataset with a sample size of 9031 snapshots. Figure 4 shows the confusion matrix for an independent dataset. The true neutron count is 5248+32=5280 and the true gamma count is 3748+3=3751. The predicted number of neutrons and gammas are 5248 and 3748 respectively. Thus, the precision for neutrons and gammas are 100% and 99%. Correspondingly the recall for neutrons and gammas are 99% and 100% respectively. The F1-score for both neutrons and gammas are 100%. Clearly, the 2D CNN-PSD model is able to classify the neutrons and gammas very accurately for independent dataset with negligible false identification. The CPU time required for the classification is 240.56 sec (4 mins) on a 2.0GHz

Intel *i*5 turbo boost system in comparison to 602.43 sec (10.04 min) for the charge integration method. Our proposed 2D CNN-PSD algorithm, unlike other ML-based PSD algorithms, cuts short the calculation and pre-processing involved in the classification of the pulses due to its ability to take snapshots. By using snapshots of the pulses, our proposed algorithm is able to classify neutrons and gamma rays based on feature extraction of the pulses and thereby reducing the computational time making it an efficient way to discriminate different types of radiations applicable for a variety of neutron detectors.

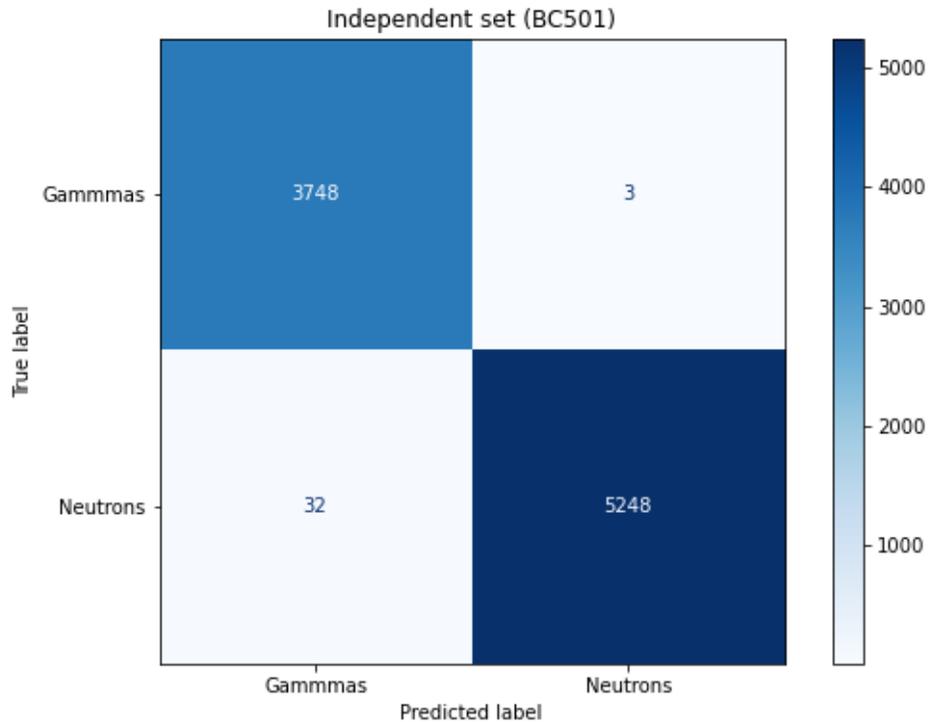

**Figure 4.** Confusion Matrix for independent dataset

**Conclusion**

In this study, we successfully showed that our proposed 2D CNN-based PSD method can discriminate neutrons from gamma rays with good accuracy. Given the proper training, the results demonstrate that the 2D CNN-PSD algorithm can distinguish between neutrons and gammas with accuracy comparable to the CI technique based on signal structures. Additionally, compared to the CI method, our suggested technique was able to accurately discriminate neutron and gamma signals from an independent data set at a faster speed. This study demonstrates that the proposed PSD method, in contrast to the CI methodology, does not necessitate computing the charge accumulated for the signals under long and short gates in order to distinguish between neutrons and gammas. Since our proposed algorithm is based on feature characterization of pulses from the snapshots, it can be applied to discriminate different types of radiation for a variety of detectors where PSD is applicable.